\documentclass[graybox]{svmult}

\usepackage{graphicx}

\usepackage{amsmath}
\usepackage{color}

\usepackage[latin1]{inputenc}
\usepackage{amsfonts}
\usepackage{tikz} 
\usetikzlibrary{matrix}
\usepackage[english]{babel}
\usepackage[latin1]{inputenc}
\usepackage[T1]{fontenc}
\usepackage{ae}

\usepackage{url}

\allowdisplaybreaks[4]

\usepackage{color}
\usepackage{tikz}
\usetikzlibrary{matrix}
\usepackage{cite}           
\usepackage{mathptmx}       
\usepackage{helvet}         
\usepackage{courier}        
\usepackage{type1cm}        
\usepackage{makeidx}         
\usepackage{graphicx}        
\usepackage{multicol}        
\usepackage[bottom]{footmisc}

\usepackage{epstopdf}

\makeindex             

\begin{document}

\title*{Analytic continuation of the kite family}
\author{Christian Bogner, Armin Schweitzer and Stefan Weinzierl}
\institute{Christian Bogner,\\ Institut f\"ur Physik, Humboldt-Universit\"at zu Berlin,
D - 10099 Berlin, Germany, \email{bogner@math.hu-berlin.de}\\
Armin Schweitzer,\\
Institut f\"ur Theoretische Physik, ETH Z\"{u}rich, 8093 Z\"{u}rich, Switzerland, \email{armin.schweitzer@phys.ethz.ch}\\
Stefan Weinzierl,\\
PRISMA Cluster of Excellence, Institut f\"ur Physik, Johannes Gutenberg-Universit\"at Mainz,
D - 55099 Mainz, Germany, \email{weinzierl@uni-mainz.de}}

%
%
\maketitle

\abstract{We consider results for the master integrals of the kite family, given
in terms of ELi-functions which are power series in the nome $q$
of an elliptic curve. The analytic continuation of these results beyond
the Euclidean region is reduced to the analytic continuation of the
two period integrals which define $q.$ We discuss the solution to
the latter problem from the perspective of the Picard-Lefschetz formula.}

%
%

\section{Introduction}

In this talk, we consider the family of Feynman integrals associated
to the kite graph, shown in fig. \ref{fig:Feynman-graphs-for} (c).
Certain master integrals of this family have recently served as interesting
showcases for the problem that multiple polylogarithms are not always
sufficient to express the coefficients of Feynman integrals in the
Laurent expansion in $\epsilon$ of dimensional regularization. Elliptic
generalizations of (multiple) polylogarithms can be used to express
these integrals instead. In \cite{AdaBogSchWei} a way to recursively
obtain the master integrals of this family to arbitrary order in $\epsilon$
was presented for the Euclidean kinematic region. This computation
and previous related work on the sunrise integral \cite{AdaBogWei1,AdaBogWei2,AdaBogWei3,AdaBogWei4}
rely crucially on properties of an underlying elliptic curve and its
periods, which were pointed out in \cite{BloVan}. The results for
the master integrals of the kite family are expressed in terms of
a class of functions defined in \cite{AdaBogWei4} as power series
in the nome $q$ of this elliptic curve. Alternative expressions in
terms of iterated integrals of modular forms were found in \cite{AdaWei}
and results for the first order of the Laurent expansion were previously
derived in \cite{RemTan2}. 

Here we focus on the analytic continuation of the results for the
kite family \cite{BogSchWei} beyond the Euclidean region. By considering
the periods of the underlying elliptic curve, we can reduce the analytic
continuation of the Feynman integrals to the question how cycles on
the elliptic curve behave under the variation of a kinematic invariant.
The answer to this question is then very simple and can be deduced
from an application of the Picard-Lefschetz formula \cite{Lef}, as
we want to emphasise with this presentation. In this way we arrive
at analytic results for the master integrals which can be evaluated
numerically at any real value of the kimematic invariant, the singular
points being the only exceptions. 

Under certain conditions, which are met in our problem, the Picard-Lefschetz
formula determines the variation undergone by integration domains
when an unintegrated variable of the integral is sent on a path in
the complex plane around a value, where a pinch singularity of the
integral occurs. It was known for a long time that at least in some
well behaved cases, the formula would apply to Feynman integrals and
predict their analytic structure. With this motivation in mind, the
theory was extended by Fotiadi, Froissart, Lascoux and especially
by Pham \cite{Fotetal,Pha1,Pha2} in the sixties, using results of
Thom \cite{Tho} and Leray \cite{Ler}. Related literature from the
sixties and seventies shows that already for rather simple Feynman
integrals a practical application of Picard-Lefschetz theory is far
from trivial. 

Since then, other methods to determine the analytic properties of
Feynman integrals have become more important. Cutkosky rules predict
the discontinuities in a handy, graphical way in terms of cut-integrals.
Furthermore, if the Feynman integral can be computed in the Euclidean
region in terms of sufficiently well-known functions such as multiple
polylogarithms, the analytic continuation to other regions can be
deduced from the analytic properties of these functions. However,
the mentioned theory framework around the Picard-Lefschetz theorem
seems to experience new attention in the recent literature on Feynman
integrals. Extended Picard-Lefschetz theory was used in a recent proof
of the Cutkosky rules in \cite{BloKre}. Furthermore, in a series
of articles \cite{Abretal1,Abretal2,Abretal3} which employs Leray's
residue theory for the definition of cut integrals, it is suggested
that the discontinuities play a crucial role in a conjectured co-product
structure on Feynman integrals, motivated from the co-product on polylogarithms.
We take these recent developments as additional motivation to emphasise
the role of homology in our application.

Our presentation is organized as follows: In the next section, we
review the family of Feynman integrals associated to the kite graph
and its underlying family of elliptic curves. In section \ref{sec:Analytic-continuation}
we reduce the problem of the analytic continuation of the master integrals
of the kite family to the question how the periods of the elliptic
curve behave under a particular variation of a kinematic parameter.
Section \ref{sec:An-application-of} discusses the latter problem
as an application of the Picard-Lefschetz formula.

\section{The kite family and its elliptic curve}

\begin{figure}
\begin{centering}
\includegraphics[scale=0.7]{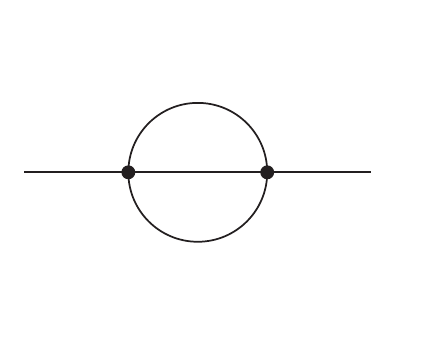}\includegraphics[scale=0.7, bb=150bp 340bp 300bp 430bp,clip]{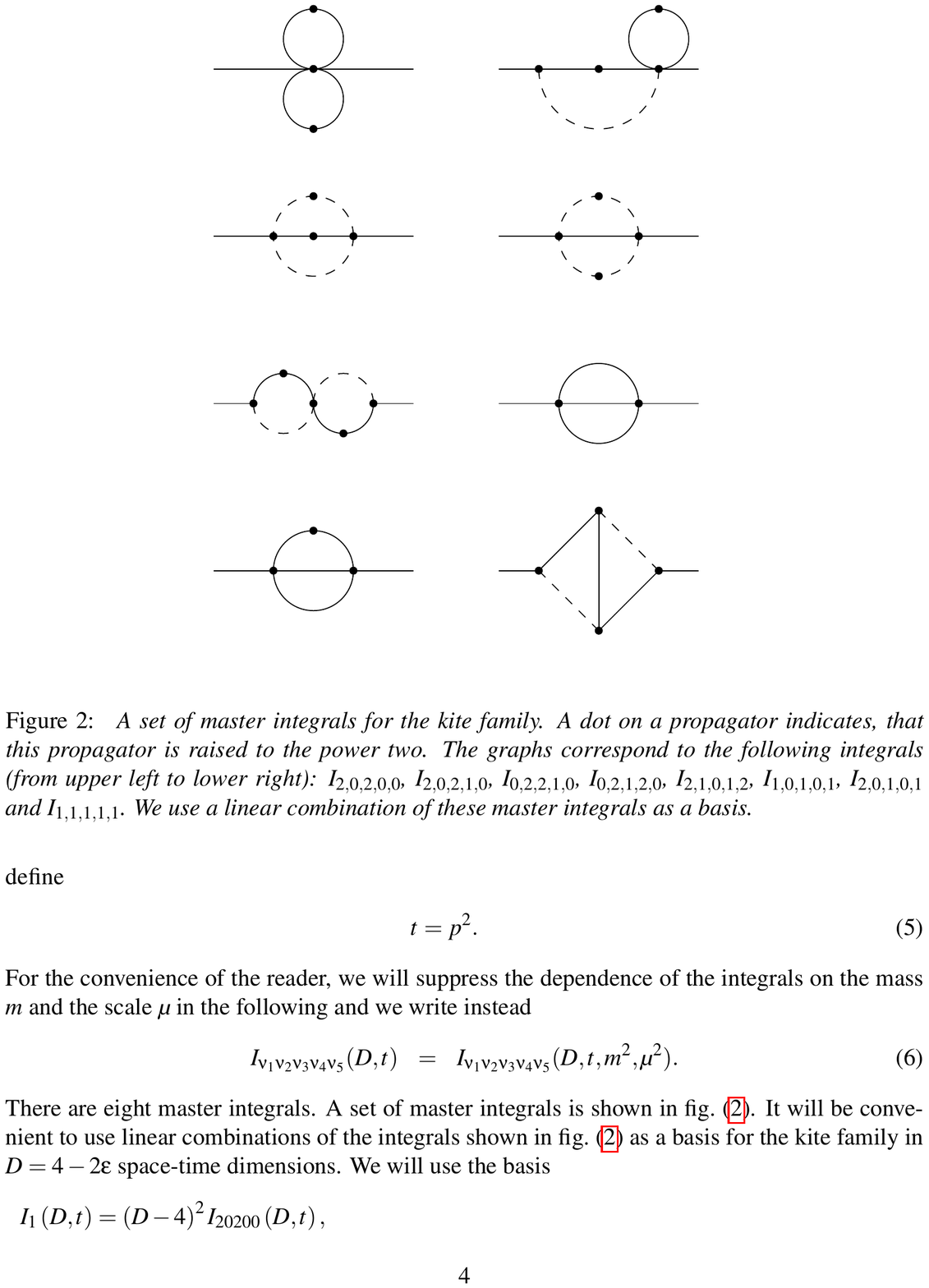}\includegraphics[scale=0.7]{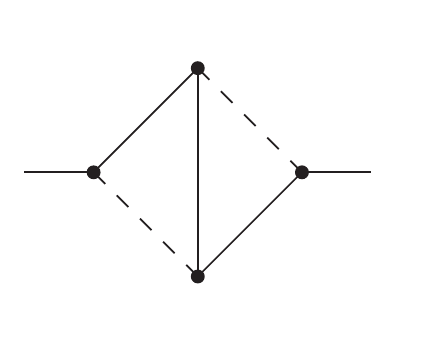}\\
(a)~~~~~~~~~~~~~~~~~~~~~~~~~~~~~~~~~~~~~~~~~(b)~~~~~~~~~~~~~~~~~~~~~~~~~~~~~~~~~~~~~~~~(c)
\par\end{centering}
\caption{The sunrise graph (a), the sunrise with one raised index (b), and
the kite graph (c).\label{fig:Feynman-graphs-for}}
\end{figure}

We consider the family of Feynman integrals associated to the kite
graph of fig. \ref{fig:Feynman-graphs-for} (c). The same particle
mass $m$ is assigned to each of the three solid internal edges while
the propagators drawn with dashed lines are massless. The graph has
one external momentum $p$ and we define $t=p^{2}.$ The integrals
of this family in $D$-dimensional Minkowski space are 
\[
I\left(\nu_{1},\nu_{2},\nu_{3},\nu_{4},\nu_{5}\right)=(-1)^{\nu}\int\frac{d^{D}l_{1}d^{D}l_{2}}{\left(i\pi^{\frac{d}{2}}\right)^{2}}\prod_{i=1}^{5}D_{i}^{-\nu_{i}}
\]
with inverse propagators $D_{1}=l_{1}^{2}-m^{2},\,D_{2}=l_{2}^{2},\,D_{3}=(l_{1}-l_{2})^{2}-m^{2},\,D_{4}=(l_{1}-p)^{2},\,D_{5}=(l_{2}-p)^{2}-m^{2}$
and $\nu=\sum_{i=1}^{5}\nu_{i}.$ The integration is over loop-momenta
$l_{1},l_{2}.$ These integrals are obviously functions of $D,t$
and $m^{2}$ which is suppressed in our notation. By integration-by-parts
reduction, the integrals of this family with $\nu_{i}\in\mathbb{Z}$
can be expressed as linear combinations of eight master integrals,
which can be chosen as $I(2,0,2,0,0),$ $I(2,0,2,1,0),$ $I(0,2,2,1,0),$
$I(0,2,1,2,0),$ $I(2,1,0,1,2),$ $I(1,0,1,0,1),$ $I(2,0,1,0,1),$
$I(1,1,1,1,1).$ The first five of these integrals can be expressed
in terms of multiple polylogarithms \cite{Gon2,Gon1}
\[
\textrm{Li}_{n_{1},...,n_{r}}(z_{1},...,z_{r})=\sum_{j_{1}>j_{2}>...>j_{r}>0}\frac{z_{1}^{j_{1}}...z_{r}^{j_{r}}}{j_{1}^{n_{1}}...j_{r}^{n_{r}}}\textrm{ for }\left|z_{i}\right|<1.
\]
The latter three integrals correspond to the graphs in fig. \ref{fig:Feynman-graphs-for}
respectively. For the computation of these inegrals, multiple polylogarithms
are not sufficient. In particular the sunrise integral $I(1,0,1,0,1)$
has been essential in recent developments to extend the classes of
functions applied in Feynman integral computations beyond multiple
polylogarithms. We refer to \cite{Abletal,AdaChaWei1,AdaChaWei2,AdaWei2,BloKerVan1,BloKerVan2,BroDuhetal1,BroDuhetal2,BroDuhetal3,Broe1,Bro2,Bro3,Bro4,ManTan,PriTan,PriTan2,RemTan3}
for some of these recent developments in quantum field theory and
string theory. 

The master integrals of the kite family can be computed by use of
the method of differential equations, deriving a system of ordinary
first-order differential equations in the variable $t.$ It was shown
in \cite{AdaBogSchWei,RemTan2} that certain changes of the basis
of master integrals simplifies the system of equations and in \cite{AdaWei2}
it was shown that by a non-algebraic change of variables, the system
can even be written in canonical form \cite{Hen2}. Results for the
master integrals were given in terms of elliptic generalizations of
(multiple) polylogarithms. In \cite{AdaBogSchWei} it was shown that
in the Euclidean region where $t<0$ the master integrals can be expressed
in terms of functions 
\begin{equation}
\textrm{ELi}_{n;m}(x;y;q)=\sum_{j=1}^{\infty}\sum_{k=1}^{\infty}\frac{x^{j}}{j^{n}}\frac{y^{k}}{k^{m}}q^{jk}=\sum_{k=1}^{\infty}\frac{y^{k}}{k^{m}}\textrm{Li}_{n}(q^{k}x),\label{eq:ELi}
\end{equation}

and multi-variable generalizations
\[
\textrm{ELi}_{n_{1},...,n_{l};m_{1},...,m_{l};2o_{1},...,2o_{l-1}}\left(x_{1},...,x_{l};y_{1},...,y_{l};q\right)
\]
\begin{equation}
=\sum_{j_{1}=1}^{\infty}...\sum_{j_{l}=1}^{\infty}\sum_{k_{1}=1}^{\infty}...\sum_{k_{l}=1}^{\infty}\frac{x_{1}^{j_{1}}}{j_{1}^{n_{1}}}...\frac{x_{l}^{j_{l}}}{j_{l}^{n_{l}}}\frac{y_{1}^{k_{1}}}{k_{1}^{m_{1}}}..\frac{y_{l}^{k_{l}}}{k_{l}^{m_{l}}}\frac{q^{j_{1}k_{1}+...+j_{l}k_{l}}}{\prod_{i=1}^{l-1}\left(j_{i}k_{i}+...+j_{l}k_{l}\right)^{o_{i}}}\label{eq:ELi multi}
\end{equation}
to all orders in $\epsilon=(4-D)/2.$ Results in terms of iterated
integrals over modular forms were derived in \cite{AdaWei}. For the
purpose of this presentation, aiming at the analytic continuation
of the results beyond the Euclidean region, the precise shape of the
results for the master integrals is not relevant. The following discussion
merely uses the fact that up to simple prefactors the results can
be expressed as power series in $q=q(t)$ which is the nome of a family
of elliptic curves, with the parameter of the family being the kinematic
invariant $t.$ 

This family of elliptic curves is derived from the sunrise integral
$I(1,0,1,0,1)$ following \cite{BloVan}. The second Symanzik polynomial
reads 
\[
\mathcal{F}=-x_{1}x_{2}x_{3}t+m^{2}\left(x_{1}+x_{2}+x_{3}\right)\left(x_{1}x_{2}+x_{2}x_{3}+x_{1}x_{3}\right).
\]
A change of variables transforms the equation $\mathcal{F}=0$ to
the Weierstrass normal form
\[
y^{2}=4\left(x-e_{1}\right)\left(x-e_{2}\right)\left(x-e_{3}\right)
\]
with the three roots 
\begin{eqnarray*}
e_{1} & = & \frac{1}{24}\left(-t^{2}+6m^{2}t+3m^{4}+3\left(m^{2}-t\right)^{\frac{3}{2}}\left(9m^{2}-t\right)^{\frac{1}{2}}\right)\\
e_{2} & = & \frac{1}{24}\left(-t^{2}+6m^{2}t+3m^{4}-3\left(m^{2}-t\right)^{\frac{3}{2}}\left(9m^{2}-t\right)^{\frac{1}{2}}\right)\\
e_{3} & = & \frac{1}{24}\left(2t^{2}-12m^{2}t-6m^{4}\right)
\end{eqnarray*}
of the cubical polynomial in $x,$ satisfying $e_{1}+e_{2}+e_{3}=0.$
The family of elliptic curves degenerates at the values $0,\,m^{2},\,9m^{2},\,\infty$
of the parameter $t.$ In the Euclidean region $t<0$ the three roots
are real and separated as $e_{1}>e_{3}>e_{2}$. Here we define the
period integrals 
\[
\psi_{1}=2\int_{e_{2}}^{e_{3}}\frac{dx}{y},\;\psi_{2}=2\int_{e_{1}}^{e_{3}}\frac{dx}{y}
\]
which evaluate to 
\[
\psi_{1}=\frac{4}{\left(m^{2}-t\right)^{\frac{3}{4}}\left(9m^{2}-t\right)^{\frac{1}{4}}}K(k),\;\psi_{2}=\frac{4i}{\left(m^{2}-t\right)^{\frac{3}{4}}\left(9m^{2}-t\right)^{\frac{1}{4}}}K(k^{\prime})
\]
with the complete elliptic integral of the first kind 
\begin{equation}
K(k)=\int_{0}^{1}\frac{dt}{\sqrt{\left(1-t^{2}\right)\left(1-k^{2}t^{2}\right)}}\label{eq:complete ellipt int}
\end{equation}
where the modulus $k$ and the complementary modulus $k^{\prime}$
are given by
\[
k=\frac{e_{3}-e_{2}}{e_{1}-e_{2}},\;k^{\prime2}=1-k^{2}=\frac{e_{1}-e_{3}}{e_{1}-e_{2}}.
\]
With these periods we introduce 
\[
\tau=\frac{\psi_{2}}{\psi_{1}},\,\,\,q=e^{i\pi\tau}.
\]
The mentioned results of \cite{AdaBogSchWei} for the eight master
integrals in the Euclidean are expressed in terms of the functions
of eqs. \ref{eq:ELi} and \ref{eq:ELi multi} with the nome $q.$
Up to simple general prefactors involving the first period $\psi_{1},$
this is their only dependence of the kinematic invariant $t.$ 

\section{Analytic continuation\label{sec:Analytic-continuation}}

\begin{figure}
\begin{centering}
\includegraphics[scale=0.8]{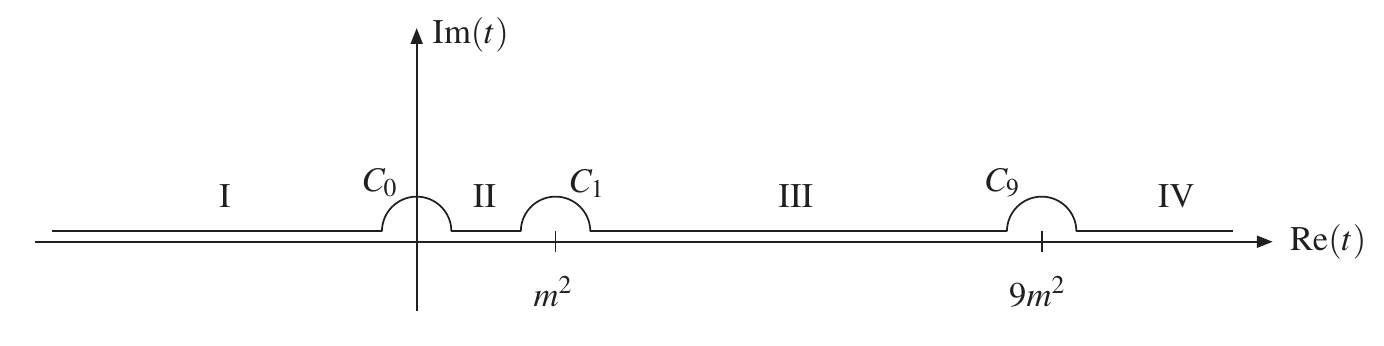}
\par\end{centering}
\caption{Variation contour in the complex $t$-plane.\label{fig:Variation-contour-in}}
\end{figure}

The previous section has shown that the analytic continuation of the
eight master integrals of the kite family can be reduced to the analytic
continuation of the two period integrals $\psi_{1},\psi_{2}.$ We
are interested in the analytic behaviour of the periods $\psi_{1},\psi_{2}$
as $t$ varies along the real axis beyond the Euclidean region. As
singular points and branch cuts of the period integrals correspond
to real values of $t,$ we consider the variation of $t$ in the complex
$t$-plane and shift the contour of this variation slightly away from
the real axis by Feynman's prescription $t\rightarrow t+i\delta.$
Here $\delta$ is small, real, positive and sent to zero in the end
for evaluations on the real axis. We choose the contour such that
it furthermore circumvents the singular points in small half circles.
Fig. \ref{fig:Variation-contour-in} shows the contour of the variation
of $t.$ 
\begin{figure}
\begin{centering}
\includegraphics[scale=0.40]{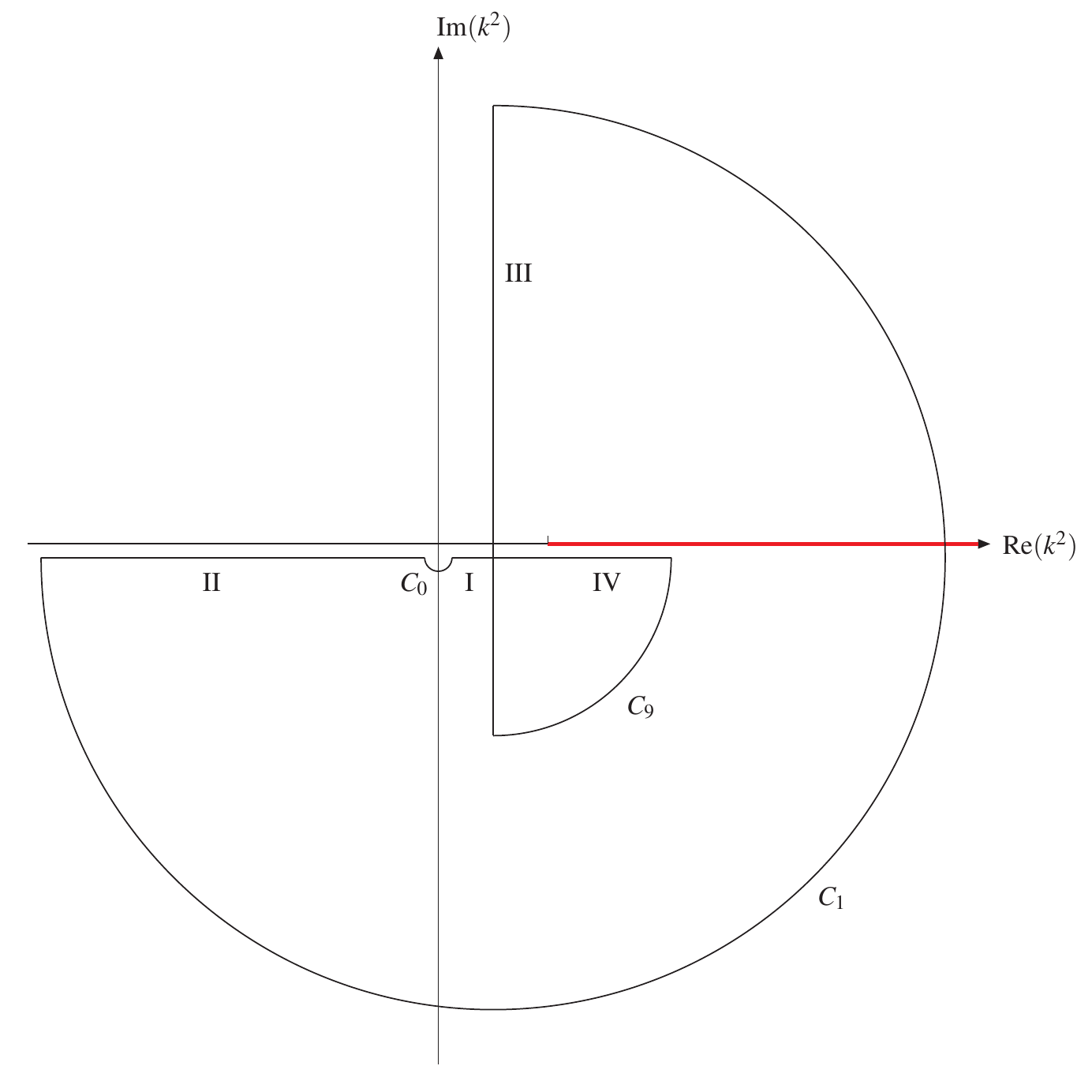}~~~\includegraphics[scale=0.40]{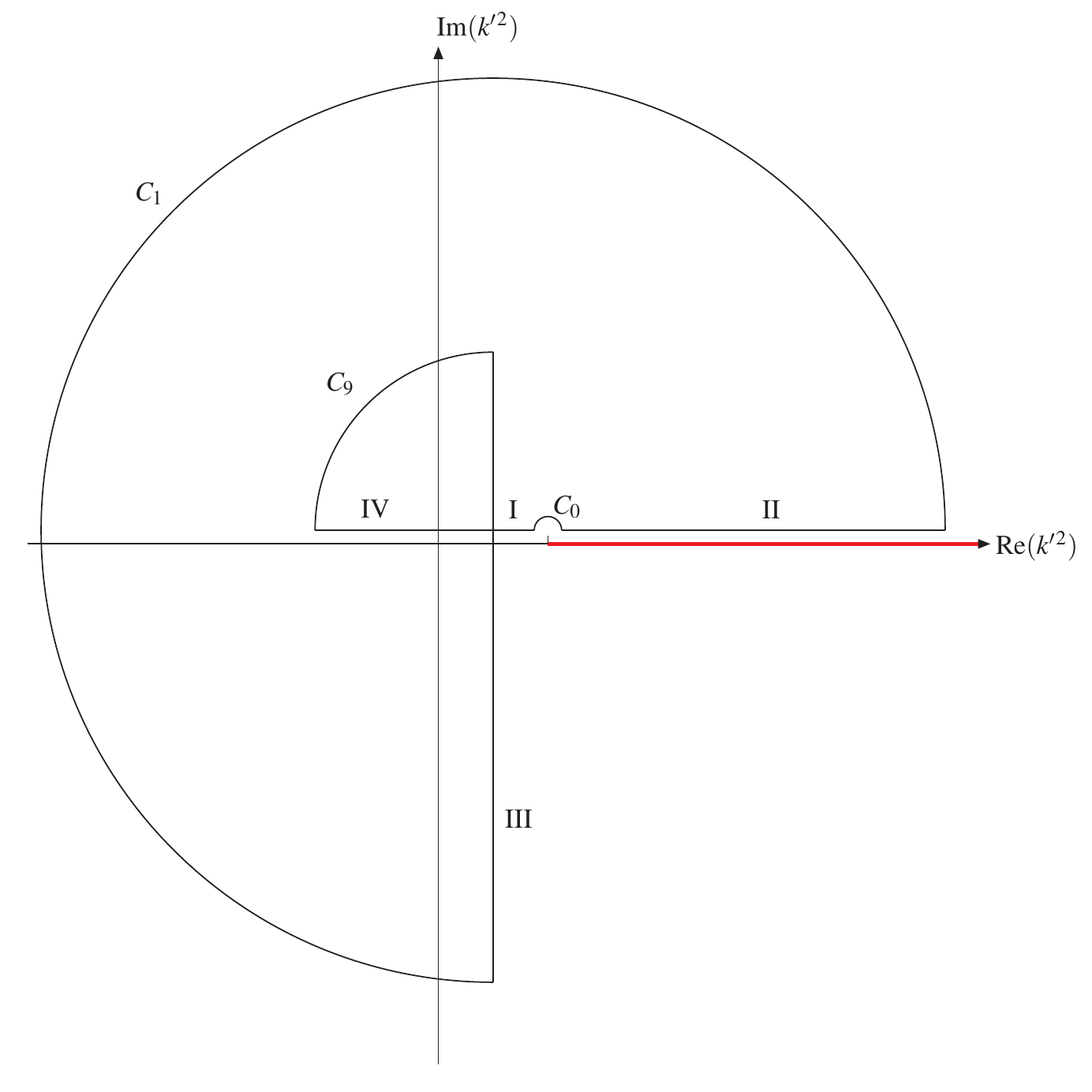}
\par\end{centering}
\caption{Variations in the complex plane of $k^{2}$ and $k^{\prime2}.$\label{fig:k^2 variations}}
\end{figure}

In order to discuss the branch cut behaviour of the periods, it is
furthermore useful to consider the complete elliptic integral of the
first kind in eq. \ref{eq:complete ellipt int} as a function of $k^{2}$
and note that it has only one branch cut $[1,\infty[$ in the complex
$k^{2}$-plane. We study the question, where along the variation of
$t$ this branch cut is crossed for the two periods. Fig. \ref{fig:k^2 variations}
shows the behaviour of $k^{2}$ and $k^{\prime2}$ as $t$ is varied
along the contour of fig. \ref{fig:Variation-contour-in}. We notice
that $k^{\prime2}$ does not cross the branch cut of the complete
elliptic integral at all. The variable $k^{2}$ crosses the branch
cut only once. This happens as $t$ is varied on the half circle $C_{1}$
around the singular point $t=m^{2}.$ Therefore it is this piece of
the contour of $t$ along which we have to study the behaviour of
the first period $\psi_{1}$ more closely.

\begin{figure}
\begin{centering}
\includegraphics[scale=0.45]{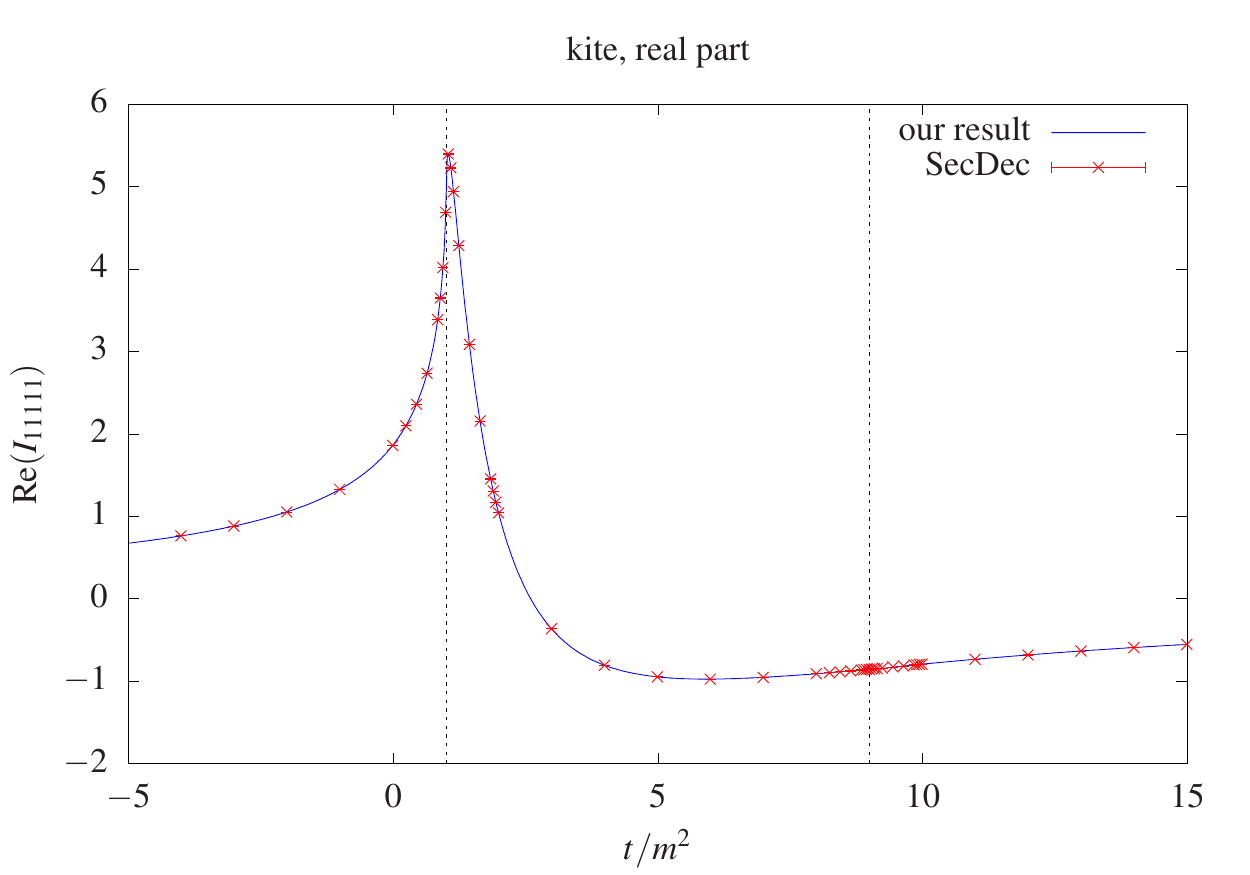}\includegraphics[scale=0.45]{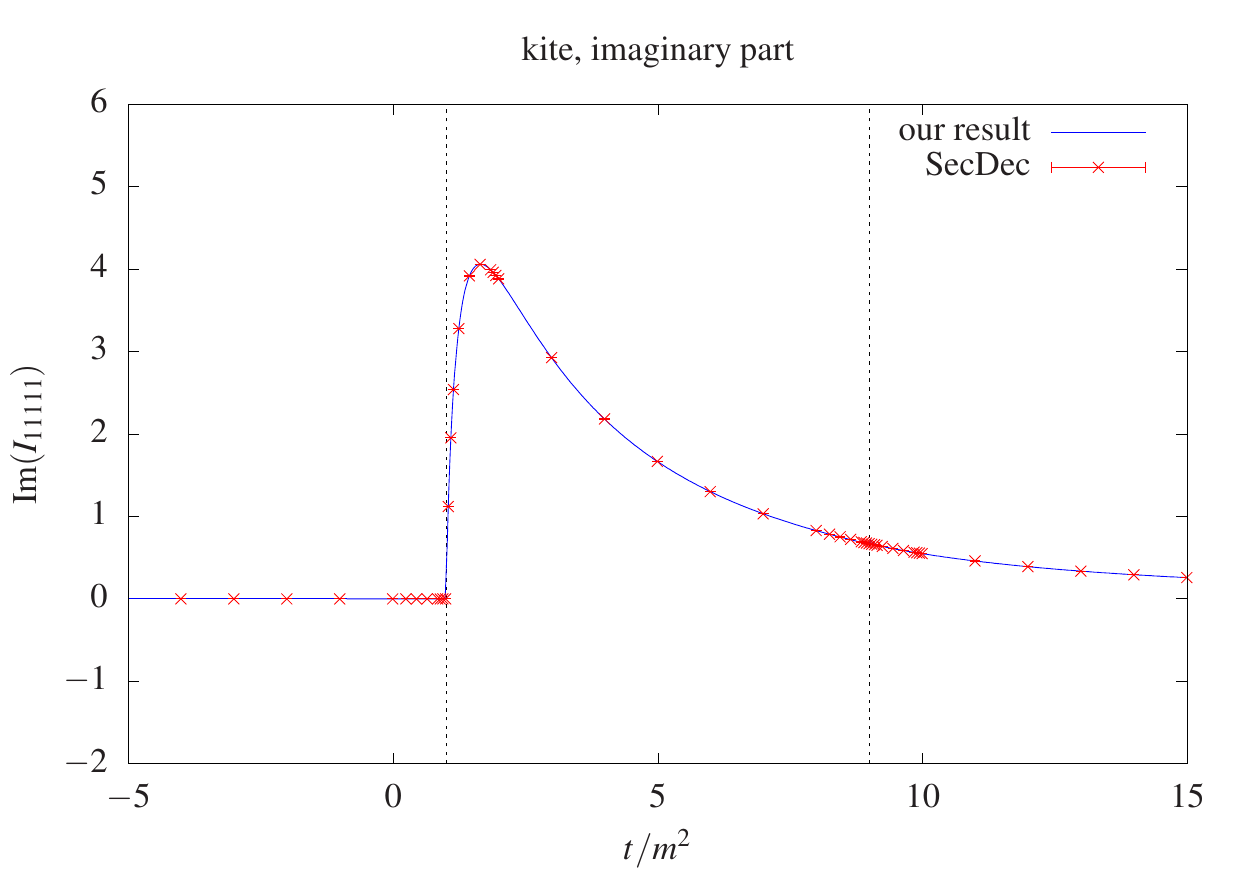}
\par\end{centering}
\caption{The real and imaginary parts of the $\epsilon^{0}$-term of the kite
integral. The dashed vertical lines indicate $t=m^{2}$ and $t=9m^{2}.$
The blue line is our analytic result and the red dots are numerical
data produced with the program SecDec \cite{Boretal}.\label{fig:numerical kite}}

\end{figure}

The three quarters of the circle which $k^{2}$ takes in fig. \ref{fig:k^2 variations}
may be deformed to a full circle for convenience. In order to study
this variation, we consider the Legendre form 
\[
y^{2}=x(x-\lambda)(x-1)
\]
of the family of elliptic curves, where $\lambda=k^{2}.$ As $t$
varies along $C_{1},$ the parameter $\lambda$ moves in a small circle
around $1.$ Equivalently, we can describe this variation by 
\[
y^{2}=x(x-e_{1}(\varphi))(x-e_{2}(\varphi))
\]

with $e_{1}(\varphi)=1-re^{i\varphi},\,\,\,\,\,e_{2}(\varphi)=1+re^{i\varphi}$
where $r$ is a small, positive, real number and $\varphi$ is an
angle whose value is 0 in the beginning and monotonously rises to
$2\pi.$ In order to observe the change of the two periods along this
variation, it is convenient to write them as integrals over cycles
$\delta_{1},\delta_{2}$ which form a basis of the first homology
group of the elliptic curve. We introduce 
\[
P_{1}(\varphi)=\int_{\delta_{1}}\frac{dx}{y},\,\,\,P_{2}(\varphi)=\int_{\delta_{2}}\frac{dx}{y},\,\,\,y=-\sqrt{x}\sqrt{x-e_{1}(\varphi)}\sqrt{x-e_{2}(\varphi)},
\]
where the cycles $\delta_{1},\delta_{2}$ are oriented such that 
\[
P_{1}(0)=2\int_{0}^{e_{1}(0)}\frac{dx}{y}=-2\int_{e_{2}(0)}^{\infty}\frac{dx}{y}\,\,\,\textrm{ and \,\,\,}P_{2}(0)=2\int_{e_{2}(0)}^{e_{1}(0)}\frac{dx}{y}
\]
with the integration contour on the right-hand side slightly shifted
by a negative imaginary part for $x.$ 
\begin{figure}
\begin{centering}
\includegraphics[scale=0.8]{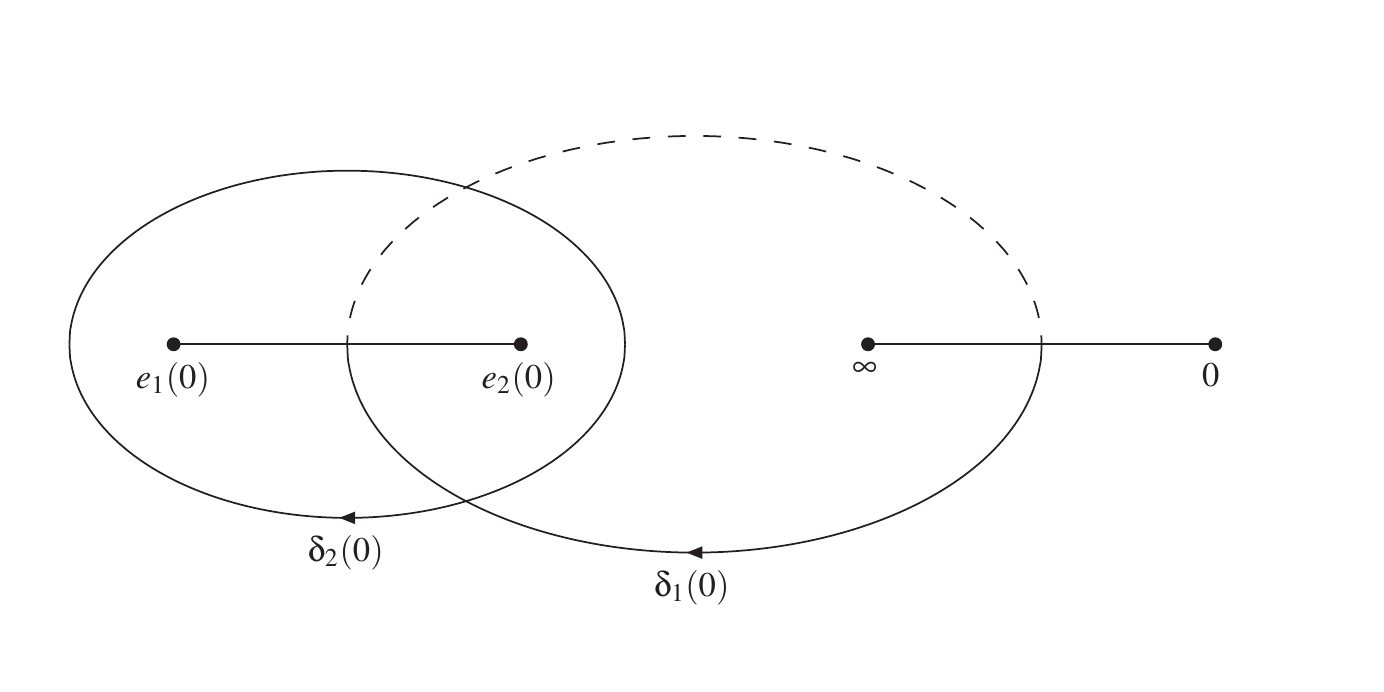}
\par\end{centering}
\caption{The cycles $\delta_{1}$ and $\delta_{2}$ before the variation.\label{fig:The-cycles-}}

\end{figure}
Fig. \ref{fig:The-cycles-} shows the cycles $\delta_{1},\delta_{2}$
on the elliptic curve. The use of dashed and straight lines indicates
that $\delta_{1}$ has two parts in two different Riemann sheets of
the elliptic curve, separated by the branch cuts. The question is:
How do the two cycles change under the mentioned variation? This will
be discussed in section \ref{sec:An-application-of}. There we will
see that $\delta_{1}$ becomes $\delta_{1}-2\delta_{2}$ while $\delta_{2}$
remains unchanged. We therefore obtain:

\[
P_{1}(2\pi)=P_{1}(0)-2P_{2}(0)\,\,\,\textrm{and}\,\,\,P_{2}(2\pi)=P_{2}(0).
\]
This is the behaviour of the periods as $t$ varies around the critical
point $t=m^{2}.$ The above discussion has shown that the behaviour
along all other pieces of the variation is trivial. We hence arrive
at the analytic continuation of the two period integrals: 
\[
\left(\begin{array}{c}
\psi_{2}(t+i\delta)\\
\psi_{1}(t+i\delta)
\end{array}\right)=\frac{4}{\left(m^{2}-t-i\delta\right)^{\frac{3}{4}}\left(9m^{2}-t-i\delta\right)^{\frac{1}{4}}}M_{t}\left(\begin{array}{c}
iK\left(k^{\prime}\left(t+i\delta\right)\right)\\
K\left(k\left(t+i\delta\right)\right)
\end{array}\right)
\]
with
\[
M_{t}=\begin{cases}
\left(\begin{array}{cc}
1 & 0\\
0 & 1
\end{array}\right) & \textrm{ for }-\infty<t<m^{2},\\
\left(\begin{array}{cc}
1 & 0\\
-2 & 1
\end{array}\right) & \textrm{ for }m^{2}<t<\infty.
\end{cases}
\]
Applying this result in terms of 
\[
q(t+i\delta)=e^{i\pi\frac{\psi_{2}(t+i\delta)}{\psi_{1}(t+i\delta)}}
\]
to the functions in eqs. \ref{eq:ELi} and \ref{eq:ELi multi}, we
obtain the analytic continuation of the results for the master integrals
of the kite family. As an example, the results for the $\epsilon^{0}$-term
for the kite integral $I(1,1,1,1,1)$ in $4-2\epsilon$ dimensions
is plotted in fig. \ref{fig:numerical kite}. 

\section{An application of the Picard-Lefschetz formula\label{sec:An-application-of}}

\begin{figure}
\begin{centering}
\includegraphics[bb=100bp 550bp 500bp 700bp,clip,scale=0.40]{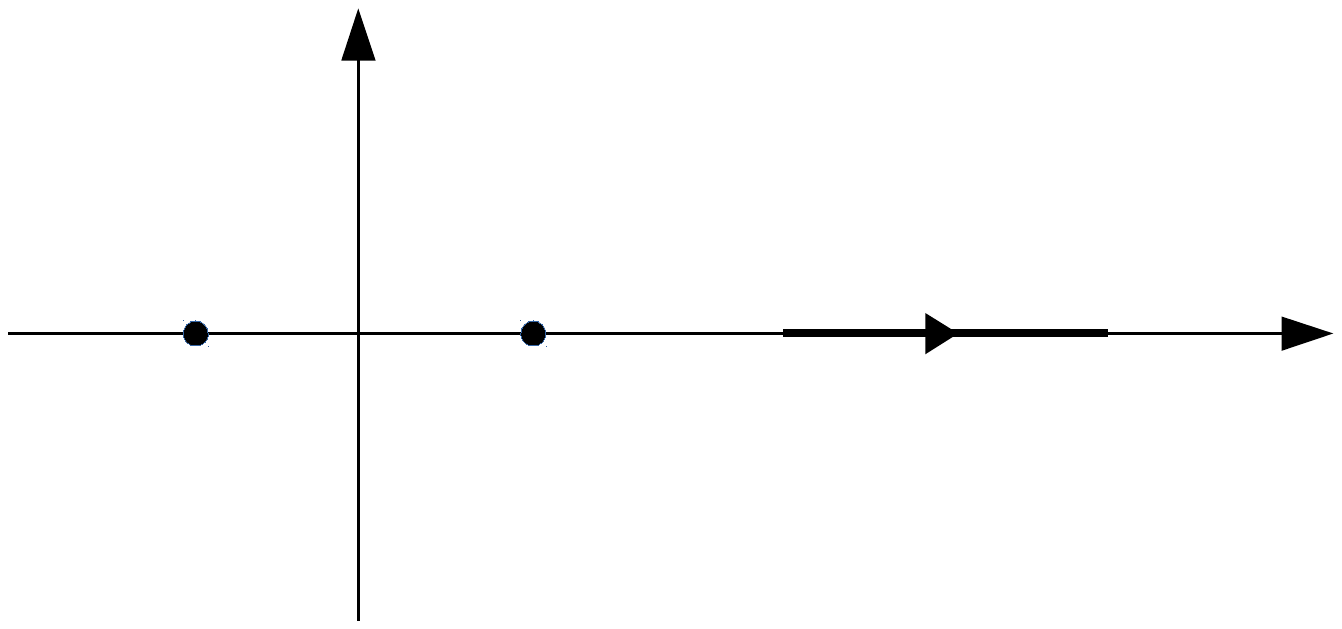}~~\includegraphics[bb=100bp 550bp 500bp 700bp,clip,scale=0.4]{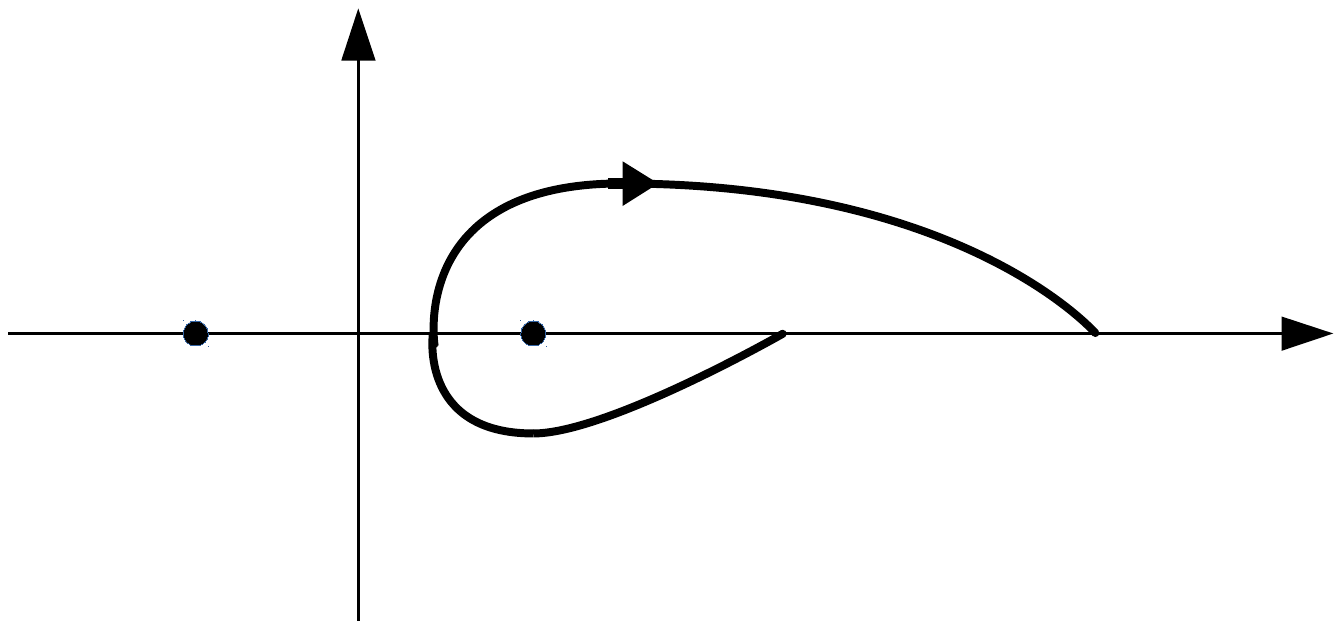}\\
(a)~~~~~~~~~~~~~~~~~~~~~~~~~~~~~~~~~~~~~~~~~~~~~~~~~~~~~~~~~~~~(b)
\par\end{centering}
\begin{centering}
\includegraphics[bb=100bp 550bp 500bp 700bp,clip,scale=0.4]{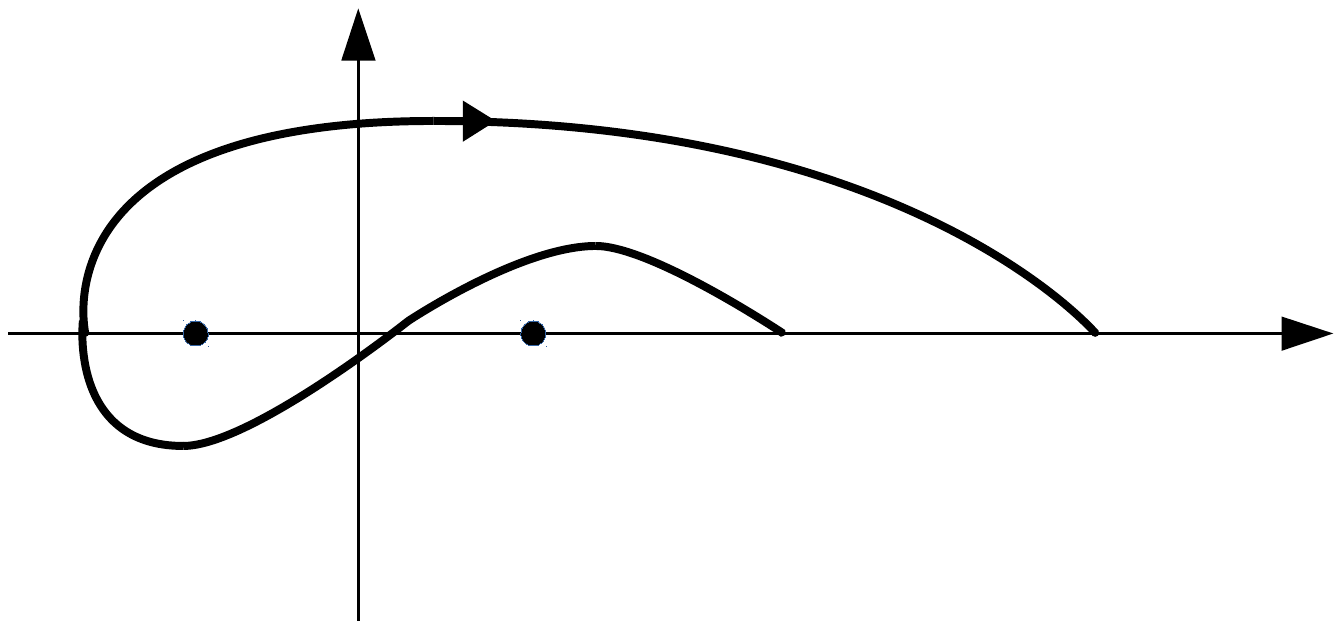}~~\includegraphics[bb=80bp 580bp 480bp 730bp,clip,scale=0.4]{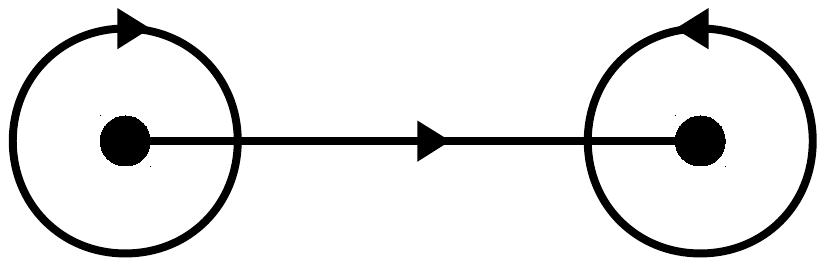}\\
(c)~~~~~~~~~~~~~~~~~~~~~~~~~~~~~~~~~~~~~~~~~~~~~~~~~~~~~~~~~~~~~~~(d)
\par\end{centering}
\begin{picture}(10,6)
\put(23,108){$e_1$}
\put(62,108){$e_2$}
\put(92,108){$a$}
\put(129,108){$b$}
\put(20,145){$\textrm{Im}(x)$}
\put(145,105){$\textrm{Re}(x)$}
\put(188,108){$e_1$}
\put(227,108){$e_2$}
\put(257,108){$a$}
\put(295,108){$b$}
\put(185,145){$\textrm{Im}(x)$}
\put(312,105){$\textrm{Re}(x)$}
\put(23,35){$e_2$}
\put(62,35){$e_1$}
\put(92,35){$a$}
\put(129,35){$b$}
\put(20,73){$\textrm{Im}(x)$}
\put(145,32){$\textrm{Re}(x)$}
\put(210,37){$e_1$}
\put(278,37){$e_2$}
\put(246,37){$s$}
\put(205,27){$c_1$}
\put(284,27){$c_2$}
\end{picture}

\caption{Contours in the complex $x$-plane.\label{fig:Contours-in-the}}
\end{figure}

Before we discuss the deformation of $\delta_{1}$ which was left
open in the previos section, let us recall the main idea of the Picard-Lefschetz
formula with the help of a classical example\footnote{Thorough introductions to Picard-Lefschetz theory can be found in
\cite{Pha2,Ebl}.} \cite{HwaTep}. We consider the integral 
\[
I(\lambda)=\int_{a}^{b}\frac{1}{x^{2}-\lambda}dx=\frac{1}{2\sqrt{\lambda}}\ln\left(\frac{\left(a+\sqrt{\lambda}\right)\left(b-\sqrt{\lambda}\right)}{\left(a-\sqrt{\lambda}\right)\left(b+\sqrt{\lambda}\right)}\right)
\]
with real $b>a>0$ depending on a complex parameter $\lambda$. We
are interested in the point $\lambda=0$ where the two singular points
$e_{1}=-\sqrt{\lambda}$ and $e_{2}=\sqrt{\lambda}$ coincide. As
long as the integration contour from $a$ to $b$ is not in between
$e_{1}$ and $e_{2},$ this contour is not trapped when the two singular
points approach each other. This is the situation of fig. \ref{fig:Contours-in-the}
(a), corresponding to the principal sheet of the logarithm. There
is no square-root singularity in this case. 

The more interesting situation is shown in fig. \ref{fig:Contours-in-the}
(b) where the integration contour is in between the points $e_{1}$
and $e_{2}$ and will be trapped for $\lambda=0.$ (This picture is
obtained after sending $\lambda$ in a small circle around $a^{2}$
in anti-clockwise direction.) The situation at $\lambda=0$ is known
as a simple pinch and it gives rise to a square-root singularity. 

Let us now send $\lambda$ in a small circle around $0$ in anti-clockwise
direction. We will call this the variation of $\lambda.$ This causes
the points $e_{1}$ and $e_{2}$ to rotate around each other in anti-clockwise
direction until they have changed positions. The result of this movement
is shown in fig. \ref{fig:Contours-in-the} (c). The integration contour
is deformed by this rotation as shown in the figure. Along the variation
of $\lambda,$ the integral $I(\lambda)$ picks up a discontinuity,
which is an integral with the same integrand and the integration contour
given by two small cycles $c_{1},c_{2}$ around $e_{1},e_{2}$ with
orientations shown in fig. \ref{fig:Contours-in-the} (d). It is easy
to see that these two cycles are in a homological sense the difference
between the integration contours of $I(\lambda)$ before and after
the variation of $\lambda$. 

It is this change of integration contours after variations around
a simple pinch which is computed in the Picard-Lefschetz formula.
The formula can be written as 
\begin{equation}
c\rightarrow c+k\cdot h,\label{eq:Picard-Lefschetz formula}
\end{equation}
where $c$ is a path or cycle, in our case the contour of integration
of $I(\lambda),$ the arrow indicates the change along the variation
of $\lambda$, $k$ is an integer and $h$ is another cycle. Both,
the integer $k$ and the cycle $h$ are determined from a so-called
vanishing cycle associated to the pinch situation. In our simple example,
the relevant vanishing cycle is the straight line $s$ oriented from
$e_{1}$ to $e_{2}$ as shown in fig. \ref{fig:Contours-in-the} (d).
This line is indeed vanishing if $\lambda$ goes to zero and it is
a relative cycle in the relative homology of the complex plane modulo
the set of points $\{e_{1},e_{2}\}.$ We may consider $s$ as an oriented
1-simplex and obtain its boundary as 
\begin{equation}
\partial s=e_{2}-e_{1}.\label{eq:partial s}
\end{equation}
The last ingredient in the construction of the cycle $h$ is the co-boundary
operator $\delta$ of Leray \cite{Ler}. The co-boundary of an $n$-dimensional
cycle can be thought of as an $(n+1)$-dimensional tube wrapped around
the cycle. In our case, we only need to construct the co-boundary
of a point, which is a small circle around this point with anti-clockwise
orientation. We obtain 
\[
h=\delta\left(\partial s\right)=c_{1}+c_{2}
\]
where the minus sign in eq. \ref{eq:partial s} is reflected in the
clockwise orientation of $c_{1}.$ 
\begin{figure}
\begin{centering}
\includegraphics[scale=0.42]{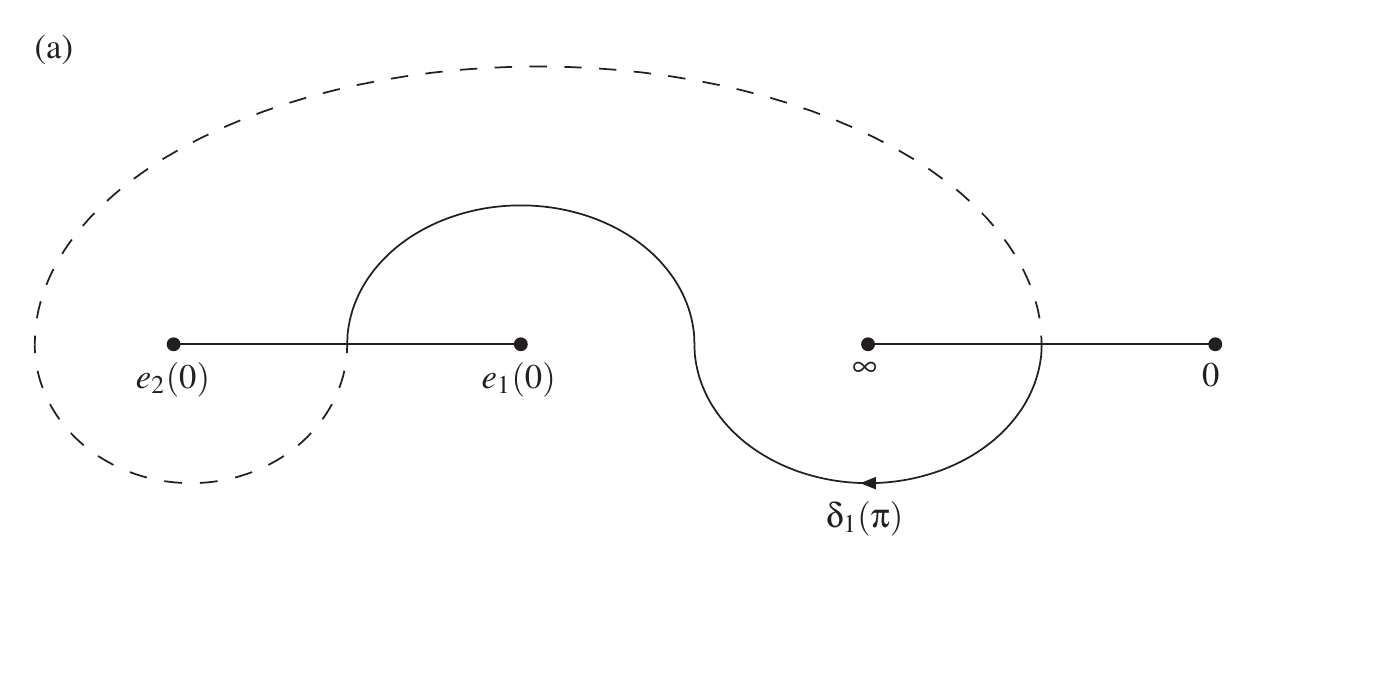}\includegraphics[scale=0.42]{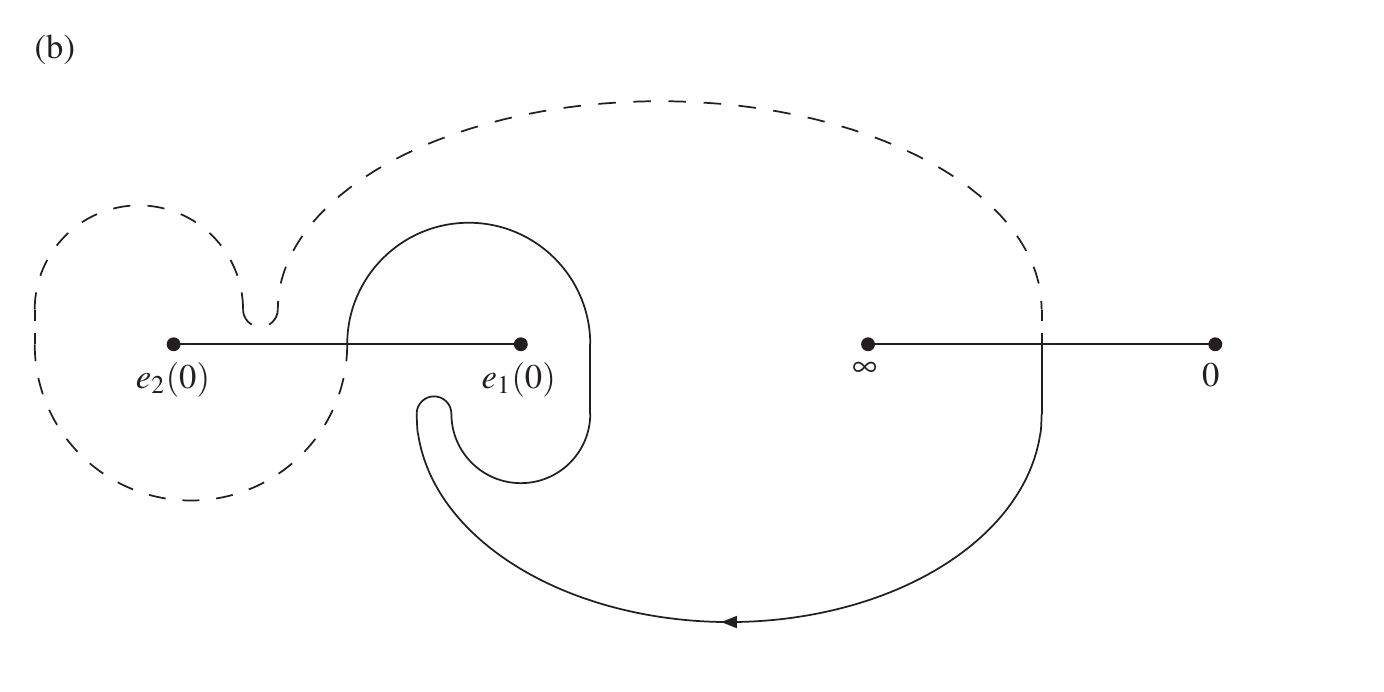}\\
\includegraphics[scale=0.42]{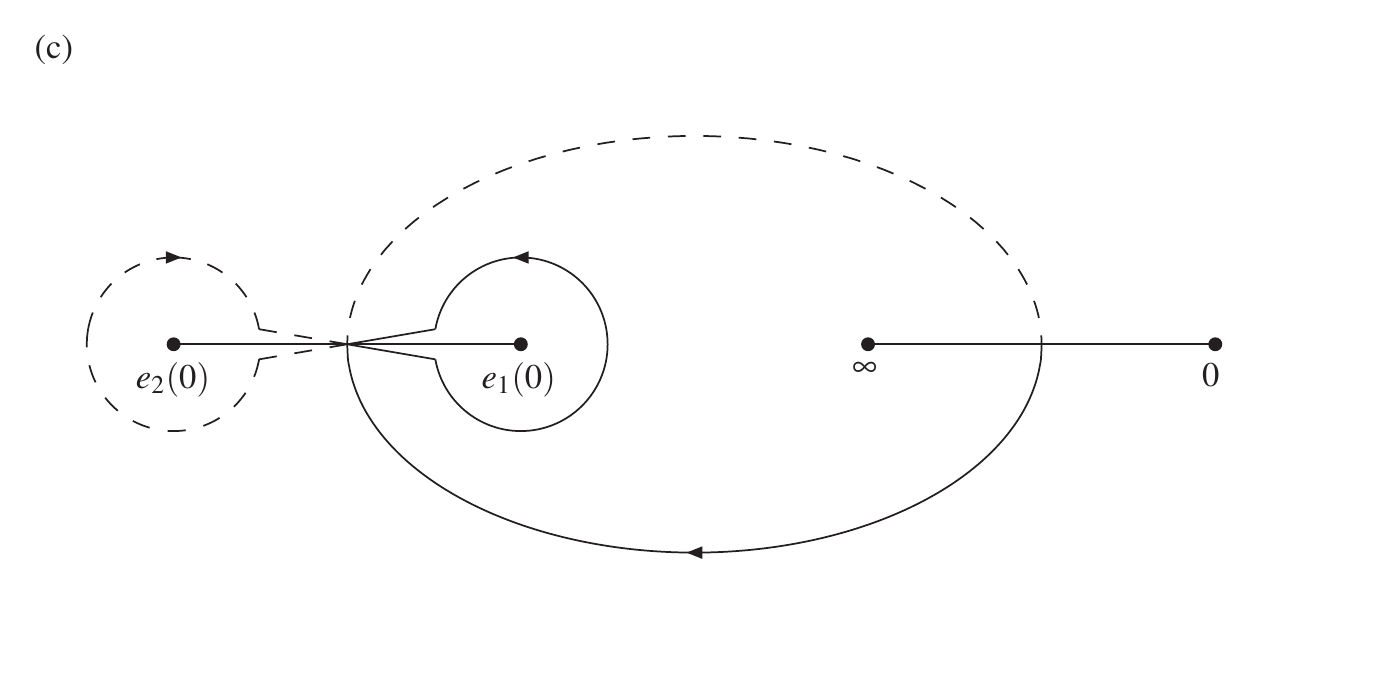}\includegraphics[scale=0.42]{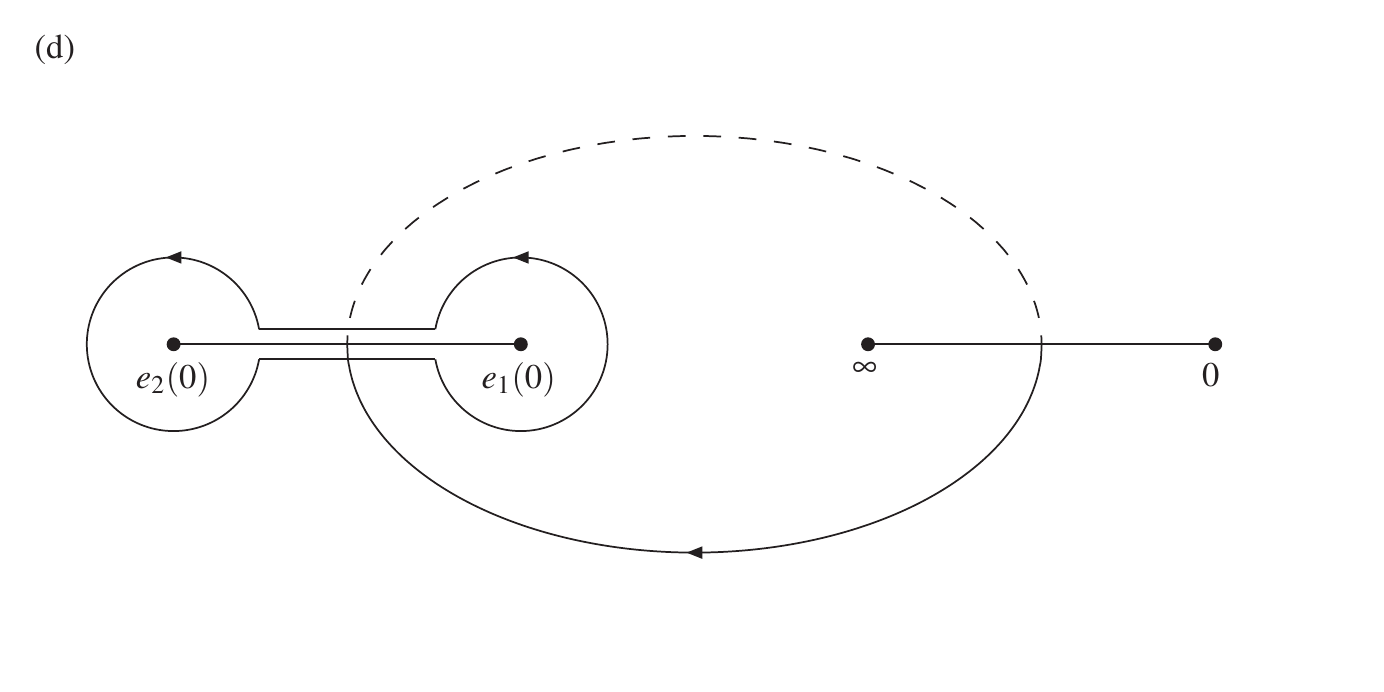}\\
\includegraphics[scale=0.42]{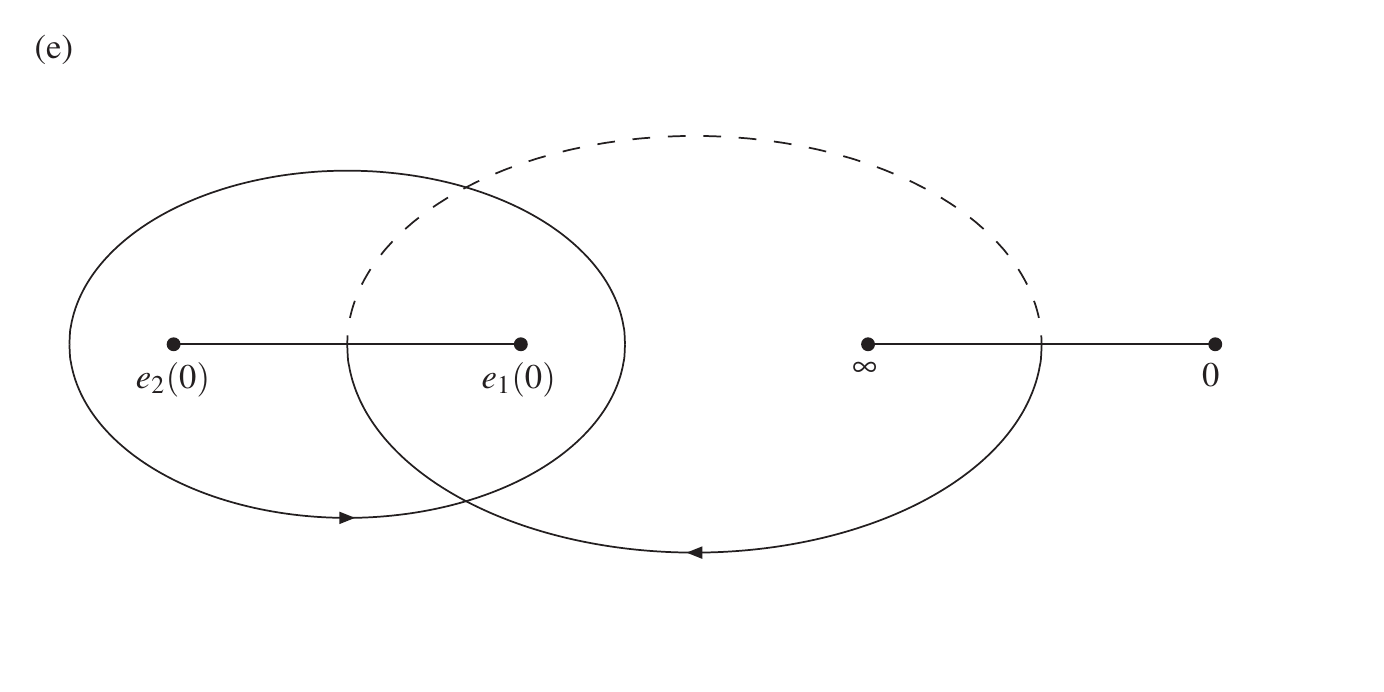}
\par\end{centering}
\caption{The deformation of $\delta_{1}$ on the elliptic curve.\label{fig:The-deformation-of}}

\end{figure}

It remains to determine the integer $k$ in the Picard-Lefschetz formula.
Up to a sign, which depends on the dimension of the problem, this
number is an intersection number or Kronecker index, depending only
on the relative orientation of the cycle $c$ and the vanishing cycle
at their intersection. In our case one simply obtains $k=1.$ In conclusion,
the Picard-Lefschetz formula predicts $c\rightarrow c+c_{1}+c_{2}$
which is precisely what we have deduced from the figures above.

We are only two steps away from the answer to the question left open
in section \ref{sec:Analytic-continuation}. On the elliptic curve,
the points $e_{1}(\lambda),$ $e_{2}(\lambda)$ coincide for $\lambda=1$
and trap the cycle $\delta_{1}$ in a simple pinch, similar to the
above example. In contrast to the warm-up example, these two points
make not half of a rotation but a full rotation around each other
as $\lambda$ is sent around the pinch point. We therefore have an
additional factor $2$ in the Picard-Lefschetz formula and obtain
\[
\delta_{1}(0)\rightarrow\delta_{1}(2\pi)=\delta_{1}(0)+2\left(c_{1}+c_{2}\right)
\]
where $c_{1}$ and $c_{2}$ are the small circles around $e_{1}$
and $e_{2}$ again. The series of snapshots in fig. \ref{fig:The-deformation-of}
shows in more detail how after half of a rotation, these circles arise
in the deformation of $\delta_{1}$ and from these pictures it is
clear, that $c_{1}$ and $c_{2}$ are located in different Riemann
sheets. In order to express the change of $\delta_{1}$ in terms of
the basis of the first homology group, $\delta_{1}(0),\delta_{2}(0),$
we may pull $c_{1}$ over to the same sheet as $c_{2}.$ This is the
step from in fig. \ref{fig:The-deformation-of} (c) to fig. \ref{fig:The-deformation-of}
(d). We see that they combine to the cycle $-\delta_{2}(0)$ and arrive
at the result 
\[
\delta_{1}(0)\rightarrow\delta_{1}(2\pi)=\delta_{1}(0)-2\delta_{2}(0)
\]
applied in section \ref{sec:Analytic-continuation}. 

We remark that this deformation on the elliptic curve is also a well-known
example. Detailed discussions with slightly different visualizations
can be found e.g. in \cite{Caretal,Zol} where the Riemann sheets,
glued together to a torus, are viewed as twisted against each other.

\vspace{1mm}\noindent

\vspace*{4mm}


\begin{thebibliography}{99.}%
\bibitem{Abletal}J. Ablinger, J. Bl\"umlein, A. De Freitas, M. van
Hoeij, E. Imamoglu, C.G. Raab, C.-S. Radu and C. Schneider, \emph{Iterated
Elliptic and Hypergeometric Integrals for Feynman Diagrams}, arXiv:1706.01299
{[}hep-th{]}.

\bibitem{Abretal1}S. Abreu, R. Britto, C. Duhr and E. Gardi, \emph{Cuts
from residues: the one-loop case}, JHEP \textbf{1706} (2017) 114,
arXiv:1702.03163 {[}hep-th{]}.

\bibitem{Abretal2}S. Abreu, R. Britto, C. Duhr and E. Gardi, \emph{Algebraic
Structure of Cut Feynman Integrals and the Diagrammatic Coaction},
Phys. Rev. Lett. \textbf{119} (2017) no.5, 051601, arXiv:1703.05064
{[}hep-th{]}.

\bibitem{Abretal3}S. Abreu, R. Britto, C. Duhr and E. Gardi, \emph{Diagrammatic
Hopf algebra of cut Feynman integrals: the one-loop case}, JHEP\textbf{
1712} (2017) 090, arXiv:1704.07931 {[}hep-th{]}.

\bibitem{AdaBogSchWei}L. Adams, C. Bogner, A. Schweitzer and S. Weinzierl,
\emph{The kite integral to all orders in terms of elliptic polylogarithms},
J. Math. Phys. \textbf{57} (2016) 122302, arXiv:1607.01571 {[}hep-ph{]}.

\bibitem{AdaBogWei1}L. Adams, C. Bogner, and S. Weinzierl, \emph{The
two-loop sunrise graph with arbitrary masses}, J. Math. Phys. \textbf{54}
(2013) 052303, arXiv:1302.7004 {[}hep-ph{]}.

\bibitem{AdaBogWei2}L. Adams, C. Bogner, and S. Weinzierl, \emph{The
two-loop sunrise graph in two space-time dimensions with arbitrary
masses in terms of elliptic dilogarithms}, J. Math. Phys. \textbf{55}
(2014) 10, 102301,arXiv:1405.5640 {[}hep-ph{]}.

\bibitem{AdaBogWei3}L. Adams, C. Bogner, and S. Weinzierl, \emph{The
two-loop sunrise integral around four space-time dimensions and generalisations
of the Clausen and Glaisher functions towards the elliptic case},
J. Math. Phys. \textbf{56} (2015) no.7, 072303, arXiv:1504.03255 {[}hep-ph{]}.

\bibitem{AdaBogWei4}L. Adams, C. Bogner, and S. Weinzierl, \emph{The
iterated structure of the all-order result for the two-loop sunrise
integral}, J. Math. Phys. \textbf{57} (2016) no.3, 032304, arXiv:1512.05630
{[}hep-ph{]}.

\bibitem{AdaChaWei1}L. Adams, E. Chaubey and S. Weinzierl, \emph{The
planar double box integral for top pair production with a closed top
loop to all orders in the dimensional regularisation parameter}, arXiv:1804.11144
{[}hep-ph{]}.

\bibitem{AdaChaWei2}L. Adams, E. Chaubey and S. Weinzierl, \emph{Analytic
results for the planar double box integral relevant to top-pair production
with a closed top loop}, arXiv:1806.04981 {[}hep-ph{]}.

\bibitem{AdaWei}L. Adams and S. Weinzierl, \emph{Feynman integrals
and iterated integrals of modular forms}, arXiv:1704.08895 {[}hep-ph{]}.

\bibitem{AdaWei2}L. Adams and S. Weinzierl, \emph{The $\epsilon$-form
of the differential equations for Feynman integrals in the elliptic
case}, Phys. Lett. \textbf{B781} (2018) 270-278, arXiv:1802.05020
{[}hep-ph{]}.

\bibitem{BloKerVan1}S. Bloch, M. Kerr and P. Vanhove, \emph{A Feynman
integral via higher normal functions}, Compos. Math. \textbf{151}
(2015) 2329-2375, arXiv:1406.2664 {[}hep-th{]}. 

\bibitem{BloKerVan2}S. Bloch, M. Kerr and P. Vanhove, \emph{Local
mirror symmetry and the sunset Feynman integral}, Adv. Theor. Math.
Phys. \textbf{21} (2017) 1373-1453, arXiv:1601.08181 {[}hep-th{]}.

\bibitem{BloKre}S. Bloch and D. Kreimer, \emph{Cutkosky Rules and
Outer Space}, arXiv:1512.01705 {[}hep-th{]}.

\bibitem{BloVan}S. Bloch and P. Vanhove, \emph{The elliptic dilogarithm
for the sunset graph}, Journal of Number Theory \textbf{148} (2015)
328\textendash 364, arXiv:1309.5865 {[}hep-th{]}. 

\bibitem{BogSchWei}C. Bogner, A. Schweitzer and S. Weinzierl, \emph{Analytic
continuation and numerical evaluation of the kite integral and the
equal mass sunrise integral}, Nucl. Phys. \textbf{B922} (2017) 528-550,
arXiv:1705.08952 {[}hep-ph{]}.

\bibitem{Boretal}S. Borowka, G. Heinrich, S.P. Jones, M. Kerner,
J. Schlenk and T. Zirke, \emph{SecDec-3.0: numerical evaluation of
multi-scale integrals beyond one loop}, Comput. Phys. Commun. \textbf{196}
(2015) 470-491, arXiv:1502.06595 {[}hep-ph{]}.

\bibitem{BroDuhetal1}J. Broedel, C. Duhr, F. Dulat and L. Tancredi,
\emph{Elliptic polylogarithms and iterated integrals on elliptic curves.
Part I: general formalism}, JHEP \textbf{1805} (2018) 093, arXiv:1712.07089
{[}hep-th{]}.

\bibitem{BroDuhetal2}J. Broedel, C. Duhr, F. Dulat and L. Tancredi,
\emph{Elliptic polylogarithms and iterated integrals on elliptic curves
II: an application to the sunrise integral}, Phys. Rev. \textbf{D97}
(2018) no.11, 116009, arXiv:1712.07095 {[}hep-ph{]}.

\bibitem{BroDuhetal3}J. Broedel, C. Duhr, F. Dulat, B. Penante and
L. Tancredi, \emph{Elliptic symbol calculus: from elliptic polylogarithms
to iterated integrals of Eisenstein series}, arXiv:1803.10256 {[}hep-th{]}.

\bibitem{Broe1}J. Broedel, C.R. Mafra, N. Matthes and O. Schlotterer,
\emph{Elliptic multiple zeta values and one-loop superstring amplitudes},
JHEP \textbf{1507} (2015) 112, arXiv:1412.5535 {[}hep-th{]}.

\bibitem{Bro2}J. Broedel, N. Matthes and O. Schlotterer, \emph{Relations
between elliptic multiple zeta values and a special derivation algebra},
J. Phys. \textbf{A49} (2016) no.15, 155203, arXiv:1507.02254 {[}hep-th{]}.

\bibitem{Bro3}J. Broedel, N. Matthes, G. Richter and O. Schlotterer,
\emph{Twisted elliptic multiple zeta values and non-planar one-loop
open-string amplitudes}, J. Phys. \textbf{A51} (2018) no.28, 285401,
arXiv:1704.03449 {[}hep-th{]}.

\bibitem{Bro4}J. Broedel, O. Schlotterer and F. Zerbini, \emph{From
elliptic multiple zeta values to modular graph functions: open and
closed strings at one loop}, arXiv:1803.00527 {[}hep-th{]}.

\bibitem{Caretal}J. Carlson, S. M\"uller-Stach and C. Peters, \emph{Period
Mappings and Period Domains}, Cambridge University Press, 2003.

\bibitem{Ebl}W. Ebeling, \emph{Funktionentheorie, Differentialtopologie
und Singularit"aten}, Vieweg Verlag, 2001.

\bibitem{Fotetal}D. Fotiadi, M. Froissart, J. Lascoux and F. Pham,
\emph{Applications of an isotopy theorem}, Topology \textbf{4} (1965),
159-191.

\bibitem{Gon2}A.B. Goncharov, \emph{Multiple polylogarithms, cyclotomy
and modular complexes}, Math. Res. Lett. \textbf{5} (1998) 497-516,
arXiv:1105.2076 {[}math.AG{]}.

\bibitem{Gon1}A.B. Goncharov, \emph{Multiple polylogarithms and mixed
Tate motives}, (2001), math.AG/0103059.

\bibitem{Hen2}J.M. Henn, \emph{Multiloop integrals in dimensional
regularization made simple}, Phys. Rev. Lett. \textbf{110} (2013)
251601, arXiv:1304.1806 {[}hep-th{]}.

\bibitem{Hok}E. D\textquoteright Hoker, M.B. Green, O. Gurdogan and
P. Vanhove, \emph{Modular Graph Functions}, Commun. Num. Theor. Phys.
\textbf{11} (2017) 165-218, arXiv:1512.06779 {[}hep-th{]}.

\bibitem{HwaTep}R.C. Hwa and V.L. Teplitz, \emph{Homology and Feynman
integrals}, W.A. Benjamin, Inc., New York, 1966.

\bibitem{Lef}S. Lefschetz, \emph{L'analysis situs et la g\'eometrie
alg\'ebrique}, Gauthier-Villars, Paris, (1924).

\bibitem{Ler}J. Leray, \emph{Le calcul diff\'erentiel et int\'egral sur
une vari\'et\'e analytique complexe (Probl\`eme de Cauchy, III)}, Bull.
Soc. Math. France, \textbf{87} (1959), 81-180.

\bibitem{ManTan}A. v. Manteuffel and L. Tancredi, \emph{A non-planar
two-loop three-point function beyond multiple polylogarithms}, JHEP
\textbf{1706} (2017) 127, arXiv:1701.05905 {[}hep-ph{]}.

\bibitem{Pha1}F. Pham, \emph{Formules de Picard-Lefschetz g\'en\'eralis\'ees
et ramification des int\'egrales}, Bull. Soc. Math. France \textbf{93},
(1965), 333-367.

\bibitem{Pha2}F. Pham, \emph{Int\'egrales Singuli\`eres}, EDP Sciences,
CNRS \'Editions, Paris, (2005).

\bibitem{PriTan}A. Primo and L. Tancredi, \emph{On the maximal cut
of Feynman integrals and the solution of their differential equations},
Nucl. Phys. \textbf{B916} (2017) 94-116, arXiv:1610.08397 {[}hep-ph{]}.

\bibitem{PriTan2}A. Primo and L. Tancredi, \emph{Maximal cuts and
differential equations for Feynman integrals. An application to the
three-loop massive banana graph}, Nucl. Phys. \textbf{B921} (2017)
316-356, arXiv:1704.05465 {[}hep-ph{]}.

\bibitem{RemTan2}E. Remiddi and L. Tancredi, \emph{Differential equations
and dispersion relations for Feynman amplitudes. The two-loop massive
sunrise and the kite integral}, Nucl. Phys. \textbf{B907} (2016) 400-444,
arXiv:1602.01481 {[}hep-ph{]}.

\bibitem{RemTan3}E. Remiddi and L. Tancredi, \emph{An Elliptic Generalization
of Multiple Polylogarithms}, Nucl. Phys. \textbf{B925} (2017) 212-25,
arXiv:1709.03622 {[}hep-ph{]}.

\bibitem{Tho}R. Thom, \emph{Les singularit\'es des applications diff\'erentiables},
Ann. Inst. Fourier \textbf{6} (1956), 43-87.

\bibitem{Zol}H. Zoladek, \emph{The Monodromy Group}, Monografie Matematyczne
\textbf{Vol. 67}, Birkh\"auser Verlag, 2006.
------------------------------------------------------------------------
\end{thebibliography}
\end{document}